\documentstyle[epsfig]{elsart}

\begin{document}

\vspace*{2cm}

\begin{center}
{\Large {\bf
$\Phi$-measure and  Disoriented chiral condensates}}
\end{center}

\bigskip

\begin{center}
{\small{
B.~Mohanty 
}}
\end{center}

\bigskip

\begin{frontmatter}

{\small{
{Institute of Physics, 751-005  Bhubaneswar, India}
}}

\begin{abstract}
Fluctuations in the ratio of neutral to charged pions arising
due to formation of disoriented chiral condensates (DCC) are discussed
using the $\Phi$-measure. The properties of the measure for 
various cases of DCC and non-DCC are discussed. The effect of detector 
efficiencies and other experimental factors are presented. Application 
of the $\Phi$-measure to simulated data, within the context of a
simple DCC model are also discussed.

\end{abstract}

\end{frontmatter}

\normalsize

In heavy-ion collisions, there is rapid expansion of collision debris
along the longitudinal direction, leading to supercooling of the
interior interaction region. This may result in  the formation of domains of
unconventionally oriented vacuum configurations as allowed by chiral
symmetry called as disoriented chiral condensates (DCC)~\cite{raj}. 
Detection of these would provide important informations on the vacuum
structure of the strong interaction and the nature of chiral phase
transition.  To have a system in which chiral symmetry is restored in the
laboratory and to be able to study the above features is one of the
main goals of ultra-relativistic heavy-ion collision experiments.

It has been predicted that DCC formation is associated with large
event-by-event fluctuations in the neutral to charged pion ratio.
The probability of the neutral pion fraction~\cite{blai,bjor}, $f$, is 
\begin{equation}
P(f) = 1/2\sqrt{f} ~~~~{\rm where}~~~
 f = N_{\pi^0}/N_{\pi},
\label{prob}
\end{equation}
$N_{\pi^0}$ and $N_{\pi}$ being the total number of neutral pions and
the total number of pions respectively.   
The corresponding distribution for non-DCC events is a gaussian with
$<f> = 1/3$. It can easily be seen that for events with DCC there is a
strong anti-correlation in the production of neutral to charged pions. 
Several theoretical calculations regarding various aspects of DCC
exists in literature, starting form the probability of DCC
production~\cite{dcc_prob} to life time of
DCC~\cite{dcc_life}. Although there is an absence of a dynamical model
of DCC, still the theoretical calculations are encouraging with respect
to formation of DCC in heavy-ion collisions.

A typical experiment, looking for DCC would consist of two detectors,
one to detect charged pions and other to detect photons from the decay 
of $\pi^{0}$'s. They must have a common $\eta$-$\phi$ overlap with
$\eta$ coverage as much as possible. From the detected hit patterns
one tries to see if there is any fluctuation in $f$, which would
indicate presence of DCC-type fluctuations. Typical event structures
would be similar to the anti-Centauro events reported by the JACEE
collaboration \cite{jacee}. Results from other cosmic ray experiments
have not ruled out the possibility of a DCC formation mechanism
\cite{augusto}. As far has accelerator based experiments are
concerned, several experiments have attempted to look for DCC  by
colliding hadrons and heavy-ions. Hadron-hadron collision experiments
like UA1~\cite{ua1_dcc}, UA5~\cite{ua5_dcc}, D0, CDF~\cite{cdf_dcc}
and MINIMAX~\cite{minimax_dcc}, having $\sqrt{s}$ from $540$ GeV
to $1.8$ TeV, have so far reported null results. Heavy-ion collision
experiments like WA98~\cite{wa98_dcc1,wa98_dcc2} and NA49~\cite{na49_dcc} 
at CERN SPS have so far put upper-limits on DCC production. In future
several experiments have planned to look for this interesting
phenomena at RHIC~\cite{star_pmd} and LHC~\cite{alice_pmd}.  

Several techniques have been developed to look for DCC. These includes 
$N_{\gamma}$ vs. $N_{ch}$ correlation~\cite{wa98_dcc2}, ``robust''
variables~\cite{minimax_dcc}  and those
based on multi-resolution analysis techniques such as discrete wavelet
transformations~\cite{wa98_dcc2}. It must be mentioned that it is
important to have the right variable or method to look for
fluctuations which are exotic. The importance of this in general has
already been emphasized by some of the recent
calculations~\cite{koch_asakawa}. One of the important observables for
looking at fluctuations is the $\Phi$-measure~\cite{phi_ebye}.  
It has been developed specifically to remove the influence of trivial
geometrical fluctuations and the effect of averaging over many particle
sources. It's utility in looking for fluctuations in transverse
momentum, azimuthal fluctuations~\cite{phi_azi} 
and studying chemical fluctuations~\cite{phi_chem,phi_chem1}
has been shown. More recently the usefulness of $\Phi$ in studies of
charge fluctuations have been demonstrated and it has been shown to
have some advantages over other techniques~\cite{phi_adv}. 
Its limitations have also been discussed~\cite{phi_lim}. In this
letter we try to see the usefulness of 
$\Phi$-measure for looking at DCC-type fluctuations.   

First we briefly recall some of the basic equations related to
$\Phi$-measure in general and then study its properties for a
DCC-type fluctuation. 
The $\Phi$-measure for a system of particles is defined as,
\begin{equation}\label{phi}
\Phi \buildrel \rm def \over = 
\sqrt{< Z^2 > \over < N >} -
\sqrt{\overline{z^2}} \;.
\end{equation}

$z$ is the single-particle variable,
$z \buildrel \rm def \over = x - \overline{x}$, 
with $\overline{x}$ being the probability (averaged over events and
particles) that a produced particle is of the sort of interest, say it
is $\pi^{0}$. One easily observes that $\overline{z} = 0$. 
While $Z$ is the event variable, which is 
a multi-particle analog of $z$, defined as 
$Z \buildrel \rm def \over = \sum_{i=1}^{N}(x_i - \overline{x})$, 
where the summation runs over particles from a given event.
By construction $< Z > = 0$, where $< ... >$ 
represents averaging over events.

As discussed in Ref.~\cite{phi_chem} one can  compute $\Phi$ for the system of
particles of two sorts, $a$ and  
$b$, e.g. $\pi^{0}$ and $\pi^{\pm}$. $x_i = 1$ when $i-$th 
particle is of the $a$ type and $x_i = 0$ otherwise. The inclusive 
average of $x$ and $x^2$ would then be given as
$$
\overline{x} = \sum_{x=0,1}xP_x = P_1 \;, 
\;\;\;\;\;\;\;\;\;\;\;\;\;\;\;\;\;
\overline{x^2} = \sum_{x=0,1}x^2P_x = P_1 \;,
$$
where $P_1$ is the probability (averaged over particles and events) 
that a produced particle is of the $a$ sort. Thus,
$$
P_1 = { < N_a > \over < N_a > +
< N_b > } \;,
$$
with $N_a$ and $N_b$ being the numbers of particles $a$ and 
$b$, respectively, in a single event. One immediately finds 
that $\overline{z} = 0 $ while
\begin{eqnarray}\label{222}
\overline{z^2} = P_1 - P_1^2 = 
{ < N_a > < N_b > \over 
< N >^2 } \;,
\end{eqnarray}

where $N = N_a + N_b$ is the multiplicity of all particles $a$ and $b$ 
in a single event.

As the event variable $Z$ is defined as $N_a - \overline{x}N$, one gets
\begin{eqnarray*}
< Z > &=& < N_a > - \overline{x}
< N > = 0 \;,\\[2mm]
< Z^2 > &=& < N_a^2 > 
- 2 \overline{x} < N_aN >
+ \overline{x}^2 < N^2 > \;.
\end{eqnarray*}
Which leads to
$$
< Z^2 > < N >^2 = 
< N_b >^2 < N_a^2 >
+< N_a >^2 < N_b^2 > 
- 2 \,< N_a > < N_b > 
< N_a N_b > \;,
$$
then
\begin{eqnarray}\label{111}
{< Z^2 > \over < N >} =
{< N_b >^2 \over < N >^3}
\big(< N_a^2 > - < N_a >^2 \big) \\ \nonumber
+{< N_a >^2 \over < N >^3} 
\big(< N_b^2 > - < N_b >^2 \big) 
\\  \nonumber
- 2{< N_a > < N_b > \over < N >^3}
\big(< N_a N_b > - < N_a > < N_b >\big)
\;.
\end{eqnarray}

The fluctuation measure $\Phi$ as given in Eqn.~(\ref{phi}) is
completely determined by  
Eqns. (\ref{222}, \ref{111}). 
Now we identify $N_{a}$ as $N_{\pi^{0}}$ and $N_{b}$ as $N_{ch} =
N_{\pi^{+}} + N_{\pi^{-}}$. If $N_{\pi} = N_{\pi^{0}} + N_{ch}$ is the 
total pion, then we can write $N_{\pi^{0}} = fN_{\pi}$ and $N_{ch} =
(1-f)N_{\pi}$. Where $f$ is the fraction of neutral pion out of total number 
of pions in a given event. With this we get,

\begin{eqnarray}\label{333}
\overline{z^2} = < f > (1 - < f >)
\end{eqnarray}

In order to calculate ${< Z^2 > \over < N_{\pi} >}$,
we consider the following equations.

\begin{eqnarray}\label{444}
{<\delta N_{\pi^{0}}^{2}>} = < f^2 > {<\delta N_{\pi}^{2}>}
+ < N_{\pi}^2 > {<\delta f^{2}>} \\ \nonumber 
+ 2 < N_{\pi} > < f  > {<\delta{N_{\pi}}\delta{f}>}
\end{eqnarray}

\begin{eqnarray}\label{555}
{<\delta N_{ch}^{2}>} = <(1-f)^2 > {<\delta N_{\pi}^{2}>}
+ < N_{\pi}^2 > {<\delta f^{2}>} \\ \nonumber
 - 2 < N_{\pi} > < (1-f)
  > {<\delta{N}\delta{f}>}
\end{eqnarray}

\begin{eqnarray}\label{666}
{<\delta{N_{\pi^{0}}}\delta{N_{ch}}>} = < f > <(1-f)
> {<\delta N_{\pi}^{2}>} - 
< N_{\pi}^2 > {<\delta f^{2}>} - \\ \nonumber 
< N_{\pi} > < f > {<\delta{N_{\pi}}\delta{f}>} +
< N_{\pi} > < (1-f) > {<\delta{N_{\pi}}\delta{f}>}
\end{eqnarray}

where, $<\delta N> = N - <N>$ and 
$<\delta N^2> = <N^{2}> - <N>^{2}$.

Using Eqn.~\ref{444}, Eqn.~\ref{555} and  Eqn.~\ref{666} in
Eqn.~\ref{111} one can easily
get the expression for  
${< Z^2 > \over < N_{\pi} >}$ as

\begin{eqnarray}\label{777}
{< Z^2 > \over < N_{\pi} >} = < N_{\pi} >
{<\delta f^{2}>} 
\end{eqnarray}
A small assumption is involved in the above derivation, 
that the fluctuations in $N_{\pi}$ is small and the term 
$\frac{<\delta {N_{\pi}}^{2}>}{<N_{\pi}>} + \frac{<{N_{\pi}}^2>}{<{N_{\pi}}>}$ 
$\sim$ $N_{\pi}$.

Using Eqn.~\ref{333} and Eqn.~\ref{777} in Eqn.~\ref{phi},
the $\Phi$-measure can be defined as,

\begin{eqnarray}\label{888}
\Phi =
\sqrt{< N_{\pi} > {<\delta f^{2}>}  } -
\sqrt{< f > (1 - < f >)} 
\end{eqnarray}

Let us now consider its properties 
within three simple models of the multiplicity distribution.

{\bf \it Non-DCC case - } 
It is known from $pp$ experiments~\cite{pp_prod} 
that the produced pions have their
charge states partitioned binomially with mean of $f$ at 1/3. The
fluctuation in $f$ is inversely proportional to the total number of 
pions given by ${<\delta f^{2}>} = <f>(1-<f>)/N_{\pi}$. Then one can easily 
see that, 

\begin{eqnarray}\label{999}
\Phi_{non-DCC} = 0
\end{eqnarray}

{\bf \it Non-DCC but $N_{\pi^{0}}$ and $N_{ch}$ are correlated - }
If $N_{\pi^{0}}$ and $N_{ch}$ are assumed to be correlated in such 
a way that there are {\it no} DCC-type fluctuations in the events. The 
neutral pion fraction, which is defined by the ratio $N_{\pi^{0}}/N_{\pi}$, 
is assumed to be strictly independent of the event multiplicity. 
Then, $f = \alpha$, where $\alpha$ is a constant smaller than 
unity.  Then ${<\delta f^{2}>} = 0$ and $\Phi$-measure is given as

\begin{equation}\label{zero-DCC}
\Phi =  - \sqrt{\alpha (1- \alpha)} \;.
\end{equation}

{\bf \it DCC case - } 
For the DCC case, the probability
distribution is given by Eqn.~(\ref{prob}).
Using this, one can easily see that $<f> = 1/3$, as in the non-DCC
case, while $<\delta{f}^2>
= 4/45$. So the $\Phi$-measure is given as

\begin{equation}\label{full-DCC}
\Phi_{DCC} =  
\sqrt{< N_{\pi} > 4/45  } -
\sqrt{2/9} 
\end{equation}

For a typical case, where the total pion multiplicity, $N_{\pi}$,
in the experiment is 300, we find for the DCC case the above
$\Phi$-measure is about 4.7.\\ This is a result where all the pions
observed are of DCC origin. However a more realistic
case~\cite{bedanga_thesis} would be to include pions from non-DCC
sources as well. 

{\bf \it Background pions - } In a given event pions can originate
both from a DCC source and 
from other non-DCC sources, even in a DCC event.
In such a case the probability distribution for the neutral
pion fraction, $P(f)$~\cite{dcc_fluct}, is given as,

\begin{eqnarray}
P(f) = \int~df_{DCC}~df_{non-DCC}~
P(f_{DCC})~P(f_{non-DCC}) \nonumber \\
 \delta(f -\beta f_{DCC}-(1-\beta)f_{non-DCC})
\label{new_prob}
\end{eqnarray}

\begin{figure}
\begin{center}
\epsfig{figure=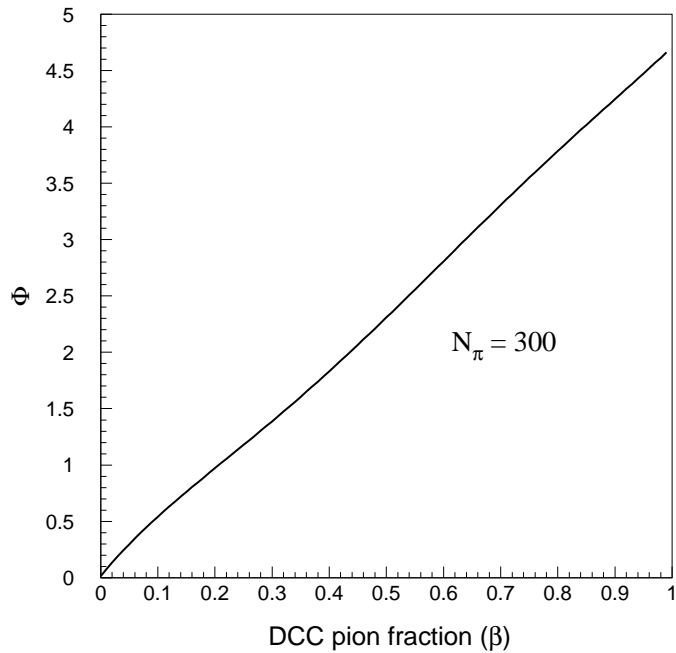,width=10cm}
\caption{\label{dcc_phi}
Variation of $\Phi$-measure as a function of the pion fraction
($\beta$) from a DCC origin. The plot is obtained for an average pion
multiplicity of 300. 
 }
\end{center}
\end{figure}

\noindent{where $\beta$ corresponds to fraction of DCC pions out of a total
number of pions $N_{\pi}$, produced in the event. $f_{DCC}$ and
$f_{non-DCC}$ correspond to the neutral pion fraction for
pions from DCC and non-DCC sources in the event, respectively.}
With the presence of non-DCC
pions, one can see, the neutral pion fraction can no longer start
from zero nor can reach the full value of unity, the range depending
upon the fraction of non-DCC pions present in the sample.
For various values of $\beta$, and total number of non-DCC pions,
$(1-\beta)N_{\pi}$, the above equation can be evaluated easily
to obtain the final $f$ distribution from which one can get the
resultant fluctuation, $<\delta{f}^2>$. Knowing $<\delta{f}^2>$
the $\Phi$-measure can be easily estimated.
The results obtained for an average total number of 300 pions
in an event, are shown in Fig.~\ref{dcc_phi}. One can see that the
value of $\Phi$-measure, decreases with decrease in
the fraction of DCC pions.

{\bf \it Multiple domains  - } It is possible that in an event
multiple domains of DCC are formed. In such a case the
total probability distribution of the neutral pion fraction is the
average value of $P(f)$ over all of the domains. This
can be written as
\begin{eqnarray}
      P_{m}(f) = \int df_{1} \cdots df_{m} \delta( f -
      \frac{f_{1}+ \cdots +f_{m}}{m}) 
 P_{1}(f_{1}) \cdots P_{m}(f_{m})
\label{mdom}
\end{eqnarray}
where, $m$ is the number of domains.
It can be shown that the resultant probability distribution approaches 
a gaussian centered at 1/3 with the
standard deviation $\sim$ $1/\sqrt{m}$.
This means ${<\delta{f}^2>}$  $\sim 1/ m$, so
that $\Phi$, reduces as the number of domains
increases in an event. 
\begin{equation}\label{multi-DCC}
\Phi_{multi-DCC} \sim
\sqrt{< N_{\pi} >/m  } -
\sqrt{2/9} 
\end{equation}
However, by carrying out this analysis
by dividing the $\eta-\phi$ phase space to appropriate bins
this effect can be reduced. 

\noindent {\bf \it Neutral pion decay. -} The neutral pions 
decay to photons ($\pi^{0}$ $\longrightarrow$ $2\gamma$), before they
reach the detectors. For such a case one can take $N_{\pi^{0}} =
2N_{\gamma}$. Then carrying out the analysis similar that done for
arriving at Eqn.~\ref{888}, one obtains

\begin{eqnarray}\label{1010}
\Phi_{\gamma} =
\sqrt{4< N_{\pi} > {<\delta f^{2}>}  } -
\sqrt{2< f > (1 - < f >)} 
\end{eqnarray}
This shows decay introduces a finite value of $\Phi$ even for non-DCC,
of the order of $0.27$

\noindent {\bf \it Detector effects -} We know that the measurement of
any observable in an experiment is affected by the
efficiency of the detector. The effect of this can be seen easily
through the following calculation.
Consider the efficiency of
detecting photons is $\epsilon_{1}$ and that for charged pions is
$\epsilon_{2}$. Then we have, $N^{exp}_{\gamma}$ = $\epsilon_{1}
N_{\gamma}$ and $N_{{ch}}^{exp}$ = $\epsilon_{2} N_{{ch}}$.
Photon multiplicity measurements are also affected by charged particle
contamination. But true $N_{\gamma}$ can be obtained from the measured
$N_{\gamma-{\mathrm like}}$ using the relation~\cite{wa98_sys}
\begin{equation}
      N_{\gamma} =
      \frac{p_\gamma}{\epsilon_{\gamma}}~N_{\gamma-{\mathrm like}}
\end{equation}
where, $\epsilon_{\gamma}$ is the photon counting efficiency,
$p_\gamma$ is the purity of the photon sample obtained from detector
simulations. For convenience one can define 
$\frac{p_\gamma}{\epsilon_{\gamma}}$ as $\epsilon_{1}$. Then following 
the simple statistical analysis and assuming that the fluctuation in 
efficiencies are independent of multiplicity (for simplicity and may
indeed be true if the analysis is carried out in narrow bins in
centrality) and fluctuations in efficiencies are small, we find
 
\begin{eqnarray}\label{1011}
\Phi_{exp} =
2\epsilon_{1} \epsilon_{2} 
\sqrt {< N_{\pi} > {<\delta f^{2}>} +
< N_{\pi} >  <f>^2 <(1-f)>^2  E }
\\ \nonumber 
- \sqrt{2 \epsilon_{1} \epsilon_{2} < f > (1 - < f >)} 
\end{eqnarray}
where, E = $\big [ \frac{<\delta \epsilon_{1}^{2}>}{{<{\epsilon_{1}}>}^2} +
\frac{<\delta \epsilon_{2}^{2}>}{{<{\epsilon_{2}}>}^2} ]$.

Let us consider an experiment where the average total detected
pion multiplicity is $300$. Taking typical values of relative
fluctuation ($\sigma/mean$) in efficiencies, to be
$\sim 3\%$~\cite{wa98_fluc}, we have
${<\delta \epsilon_{1}>}^2/{{<{\epsilon_{1}}>}^2}$ and
${<\delta \epsilon_{2}>}^2/{{<{\epsilon_{2}}>}^2}$ both to 
be about $0.0009$. Together they introduce an error
$0.0027$. Then the term 
${< N_{\pi} >  <f>^2 <(1-f)>^2 \big [ \frac{<\delta
  \epsilon_{1}^{2}>}{{<{\epsilon_{1}}>}^2} + \frac{<\delta
  \epsilon_{2}^{2}>}{{<{\epsilon_{2}}>}^2} ]}$ $\sim$ 0.04 
which is very small compared to $< N_{\pi} > {<\delta f^{2}>}$
so may be for simplicity neglected. Hence $\Phi_{exp}$ can be now
written as,

\begin{eqnarray}\label{1012}
\Phi_{exp} =
2\epsilon_{1} \epsilon_{2} \sqrt{< N_{\pi} > {<\delta f^{2}>}}
- \sqrt{2 \epsilon_{1} \epsilon_{2} < f > (1 - < f >)} 
\end{eqnarray}
So knowing the efficiency and purity of the detected multiplicity
sample, one can estimate the value of $\Phi_{exp}$.

\noindent {\bf \it Application to simulated data -} 
We have discussed above many factors that possibly affect the
detection of DCC-type fluctuations in data. Now lets apply the
$\Phi$-measure to simulated data and observe the sensitivity of the measure
to detection of DCC-type fluctuations. Another advantage of applying
it to simulated data, is that it can form a guidance for application
of this to actual data in future.
Simulated events were generated using VENUS 4.12 \cite{venus} event
generator with default parameters. 
15K VENUS events with impact parameter less than $3~fm$ was generated
for this study. We have compared the observed effect
to simulation results based on a simple DCC model \cite{wa98_dcc1,plbeshape}.
We assume the formation of a single DCC domain of a given size
lying within the limited coverage ($\eta$ = 3-4, and full azimuthal) 
of a hypothetical detector system,
consisting of a photon multiplicity detector and a charged particle 
multiplicity detector.
In this model the output of the VENUS~4.12 event generator
has been suitably
modified to accommodate such a domain, characterized by percentage of
pions being DCC type. To introduce DCC in a certain fraction of pions out of
the total pions in an event, the charge of the pions is interchanged
pairwise ($\pi^{+}\pi^{-} \leftrightarrow \pi^{0}\pi^{0}$),
according to the DCC probability as given by Eqn. (1).
After allowing the $\pi^0$s to decay, all
the particles are passed through the realistic detector 
responses. The efficiency of charged particle detection was taken to
be $90\% \pm 5\%$~\cite{spmd_nim}. The photon counting efficiency and 
purity of photon sample was taken to be same and $70\% \pm
5\%$~\cite{wa98_sys}. The average number of photons detected are $372$
and the average number of charged particles detected are $456$.

\begin{figure}
\begin{center}
\epsfig{figure=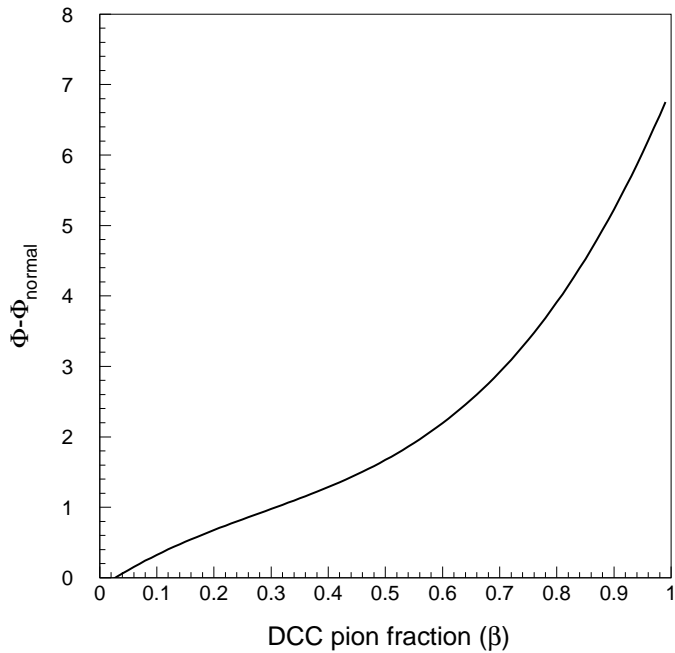,width=10cm}
\caption{\label{dcc_phi1}
Variation of $\Phi - \Phi_{normal}$ as a function of the pion fraction
($\beta$) from a DCC origin, obtained from simulated data using VENUS.
 }
\end{center}
\end{figure}

First we study the effect of number of pions being DCC-type.
As discussed above and shown in Fig.~\ref{dcc_phi}, the value
of $\Phi$ decreases as the number of non-DCC pions in a given event
increases. Here we study the same in a realistic scenario using
simulated data. We introduce DCC for different fractions of pions
being DCC-type ($\beta$) in an event. This is done following the
method discussed above. Then we calculate the quantity, $\Phi -
\Phi_{normal}$, Where $\Phi$ is the measure of fluctuation for a given 
DCC pion fraction ($\beta$) and $\Phi_{normal}$ is the measure of
fluctuation for normal or non-DCC type  events (detector and decay
effects included). The variation of $\Phi - \Phi_{normal}$ as a
function of $\beta$ is shown in Fig.~\ref{dcc_phi1}. 
From the Fig.~\ref{dcc_phi1} we find that the
fluctuations decreases with decrease with DCC pion fraction, which is
consistent with the theoretical calculation shown in Fig.~\ref{dcc_phi}.
The statistical error on $\Phi$ measure was calculated by taking
different number of non-DCC type events (1000, 2000, $\cdots$,10000)
then finding the maximum variation  in $\Phi$ value for these sets. It
was calculated to be $0.006$. From this we find that such a typical
experiment as discussed in the generating simulated data, will be able 
to detect DCC-type fluctuation, if the number of pions in an event
being DCC-type is above $\sim$ $3\%$.

\begin{figure}
\begin{center}
\epsfig{figure=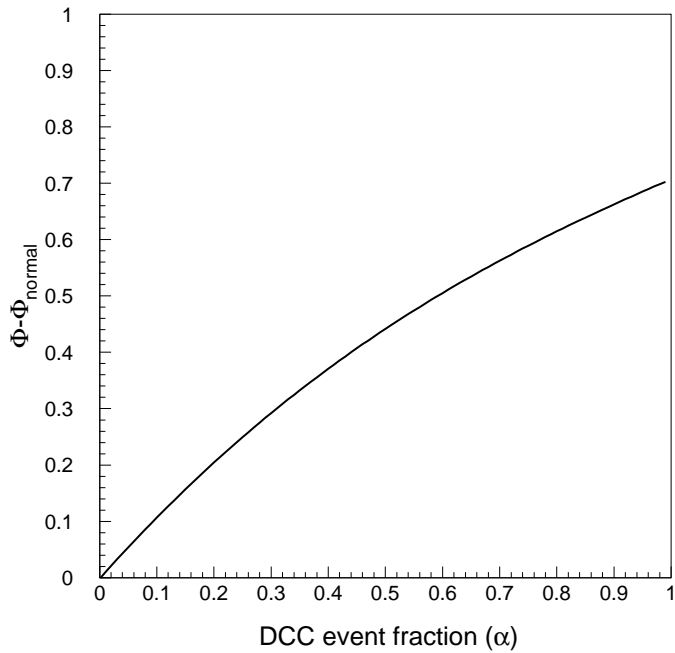,width=10cm}
\caption{\label{dcc_phi2}
Variation of $\Phi - \Phi_{normal}$  as a function of the DCC event fraction
($\alpha$) for a fixed DCC pion fraction ($\beta$ = 0.25), obtained from
simulated data using VENUS.
 }
\end{center}
\end{figure}
Secondly, we know that all events in a given heavy-ion reaction cannot 
be of DCC type. The effect of this has been studied for a fixed number 
of pions being DCC-type. For simplicity, we have assumed that for DCC
type events the percentage of DCC-type pions out of the total number of
pions is $25\%$. Then we have varied the number of events being
DCC-type ($\alpha$) out of the total number of events in an given
ensemble of events, to study the effect of DCC event fraction
($\alpha$) on $\Phi$-measure. The results have been presented as $\Phi
- \Phi_{normal}$ vs. $\alpha$ and are shown in Fig.~\ref{dcc_phi2}. 
We find that the value of $\Phi - \Phi_{normal}$ decreases with
decrease in $\alpha$, which is along expected lines. Keeping in mind
the above mentioned statistical error on $\Phi$-measure, we find that
for DCC pion fraction ($\beta$) of 0.25, the measure is sensitive to  
more than $1\%$ of events being DCC-type out of a given ensemble of
events. Similar calculations can be done for different values of
$\beta$, but not discussed here, as the aim here is to demonstrate the 
utility of this widely used observable for DCC-type of studies. 

\noindent {\bf \it Summary -} We have discussed the utility and
sensitivity of $\Phi$-measure for looking at DCC-type fluctuations. We
have studied the properties of $\Phi$-measure for three different
simple models of multiplicity distributions. Then we discussed  the
effect of various factors affecting the DCC-type fluctuations on
$\Phi$-measure. These include the existence of background or non-DCC
type pions in addition to those  from DCC origin in an event, possible
existence of multiple domains of DCC, decay of neutral pion to
photons and the detector effects like  efficiency and purity. We have
then discussed extensively the variation of $\Phi$ with fraction of
pions being DCC type ($\beta$) and fraction of events being DCC type
($\alpha$), within the framework  of a simple DCC model. From the study
using simulated data from VENUS and incorporating realistic detector
effects, we found that for a typical example, such a measure is
sensitive to DCC-type fluctuations with $\beta$ greater than $0.03$
and for a $\beta = 0.25$, it is sensitive to DCC-type fluctuation
with $\alpha$ greater than $0.01$. With this study we have
demonstrated  that $\Phi$-measure is a sensitive observable to look
for DCC-type fluctuations, while it preserves its other properties,
which has made it a powerful measure in event-by-event studies.

{\bf Acknowledgments} \\
We would like to thank A.K.~Dubey of Institute of Physics, 
for some useful discussions during the initial stages of this work.

\end{document}